\documentclass{article}
\usepackage{breckenr2005}

\frompage{000} \topage{000}                                              
\def\rightmark{Determining the Density Dependence of Nuclear Symmetry 
Energy}
\def\leftmark{Bao-An Li et al.}
\authors{
{Bao-An Li$^1$, Lie-Wen Chen$^{2}$, Che Ming Ko$^{3}$, 
Gao-Chan Yong$^{4}$ and Wei Zuo$^{4}$ }\\[2.812mm]
{\normalsize
\hspace*{-8pt}$^1$ Department of Chemistry and Physics, Arkansas State 
University,\\ P.O. Box 419, State University, AR 72467-0419, USA \\[0.2ex] 
\hspace*{-8pt}$^2$ Institute of Theoretical Physics, Shanghai Jiao
Tong University,\\ Shanghai 200240, P.R. China\\[0.2ex]
\hspace*{-8pt}$^3$ Cyclotron Institute and Department of Physics, 
Texas A\&M University,\\ 
College Station, TX 77843-3366, USA\\[0.2ex]
\hspace*{-8pt}$^4$ Institute of Modern Physics, Lanzhou 730000, and 
Graduate School, Beijing 100039, Chinese Academy of Science, P.R. China
}}
\abstract{The latest development in determining the density dependence
of the nuclear symmetry energy using heavy-ion collisions is
reviewed. Within the IBUU04 version of an isospin- and
momentum-dependent transport model using a modified Gogny effective 
interaction, recent experimental data from NSCL/MSU on isospin diffusion are 
found to be consistent with a nuclear symmetry energy of 
$E_{sym}(\rho )\approx 31.6(\rho /\rho_{0})^{1.05}$ at subnormal densities. 
Predictions on several observables sensitive to the density dependence
of the symmetry energy at supranormal densities accessible at GSI and
the planned Rare Isotope Accelerator (RIA) are also made.}

\begin{document}

\title{Progress Towards Determining the Density Dependence of the Nuclear
Symmetry Energy Using Heavy-Ion Reactions}
\author{}
\maketitle

\setcounter{page}{1}

\section{Introduction}

\label{intro} The Equation of State (EOS) of isospin asymmetric nuclear
matter can be written within the well-known parabolic approximation,
which has been verified by all many-body theories, as 
\begin{equation}  \label{ieos}
E(\rho ,\delta )=E(\rho ,\delta =0)+E_{sym}(\rho )\delta ^{2}+\mathcal{O}%
(\delta^4),
\end{equation}
where $\delta\equiv(\rho_{n}-\rho _{p})/(\rho _{p}+\rho _{n})$ is the
isospin asymmetry and $E_{sym}(\rho)$ is the density-dependent nuclear
symmetry energy. The latter is very important for many interesting
astrophysical problems\cite{lat01}, the structure of radioactive nuclei\cite%
{brown,stone} and heavy-ion reactions\cite{ireview98,ibook01,dan02,ditoro}.
Unfortunately, the density dependence of symmetry energy $E_{sym}(\rho)$,
especially at supranormal densities, is still poorly known. Predictions
based on various many-body theories diverge widely at both low and high
densities. In fact, even the sign of the symmetry energy above
$3\rho_0$ remains uncertain\cite{bom1}. Fortunately, heavy-ion
reactions, especially those induced by radioactive beams, provide a 
unique opportunity to pin down the density dependence of nuclear 
symmetry energy in terrestrial laboratories. Significant progress in 
determining the symmetry energy at subnormal densities has been made 
recently both experimentally and theoretically\cite{betty04,chen05}.  
High energy radioactive beams to be available at GSI and RIA will
allow us to determine the symmetry energy at supranormal densities. In 
this talk, we highlight the recent most exciting progress in
determining the symmetry energy at subnormal densities and
present our predictions on several most sensitive probes of the symmetry
energy at supranormal densities.

\section{An isospin- and momentum-dependent transport model for nuclear
reactions induced by radioactive beams}

Crucial to the extraction of critical information about the $E_{\mathrm{sym}%
}(\rho )$ is to compare experimental data with transport model calculations.
We outline here the major ingredients of the version IBUU04 of an isospin-
and momentum-dependent transport model for nuclear reactions induced by
radioactive beams\cite{lidas03}. The single nucleon potential is one of the
most important inputs to all transport models. Both the isovector (symmetry
potential) and isoscalar parts of this potential should be momentum
dependent due to the non-locality of strong interactions and the Pauli
exchange effects in many-fermion systems. In the IBUU04, we use a single
nucleon potential derived from the Hartree-Fock approximation using a
modified Gogny effective interaction (MDI)\cite{das03}, i.e.,   
\begin{eqnarray}
U(\rho ,\delta ,\vec{p},\tau ,x) &=&A_{u}(x)\frac{\rho _{\tau ^{\prime }}}{%
\rho _{0}}+A_{l}(x)\frac{\rho _{\tau }}{\rho _{0}}  \nonumber  \label{mdi} \\
&+&B(\frac{\rho }{\rho _{0}})^{\sigma }(1-x\delta ^{2})-8\tau x\frac{B}{%
\sigma +1}\frac{\rho ^{\sigma -1}}{\rho _{0}^{\sigma }}\delta \rho _{\tau
^{\prime }}  \nonumber \\
&+&\frac{2C_{\tau ,\tau }}{\rho _{0}}\int d^{3}p^{\prime }\frac{f_{\tau }(%
\vec{r},\vec{p}^{\prime })}{1+(\vec{p}-\vec{p}^{\prime })^{2}/\Lambda ^{2}} 
\nonumber \\
&+&\frac{2C_{\tau ,\tau ^{\prime }}}{\rho _{0}}\int d^{3}p^{\prime }\frac{%
f_{\tau ^{\prime }}(\vec{r},\vec{p}^{\prime })}{1+(\vec{p}-\vec{p}^{\prime
})^{2}/\Lambda ^{2}}.
\end{eqnarray}%
In the above $\tau =1/2$ ($-1/2$) for neutrons (protons) and $\tau \neq \tau
^{\prime }$; $\sigma =4/3$; $f_{\tau }(\vec{r},\vec{p})$ is the phase space
distribution function at coordinate $\vec{r}$ and momentum $\vec{p}$. The
parameters $A_{u}(x),A_{l}(x),B,C_{\tau ,\tau },C_{\tau ,\tau ^{\prime }}$
and $\Lambda $ were obtained by fitting the momentum-dependence of the $%
U(\rho ,\delta ,\vec{p},\tau ,x)$ to that predicted by the Gogny
Hartree-Fock and/or the Brueckner-Hartree-Fock calculations, the saturation
properties of symmetric nuclear matter and the symmetry energy of 30 MeV at
normal nuclear matter density $\rho _{0}=0.16$ fm$^{-3}$\cite{das03}. The
incompressibility $K_{0}$ of symmetric nuclear matter at $\rho _{0}$ is set
to be 211 MeV. The parameters $A_{u}(x)$ and $A_{l}(x)$ depend on the $x$
parameter according to $A_{u}(x)=-95.98-x\frac{2B}{\sigma +1}$ and $%
A_{l}(x)=-120.57+x\frac{2B}{\sigma +1}$. The parameter $x$ can be adjusted
to mimic predictions on the $E_{sym}(\rho )$ by microscopic and/or
phenomenological many-body theories. 
\begin{figure}[tbh]
\insertplot{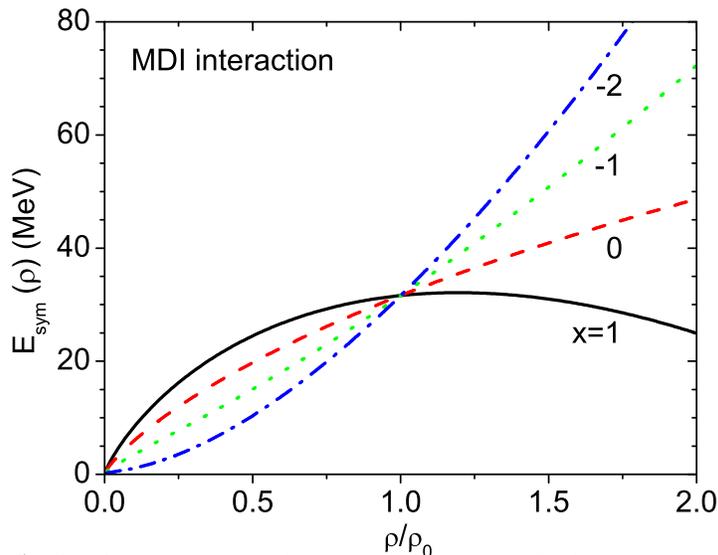} \vspace*{-1cm}
\caption{{\protect\small Density dependence of the symmetry energy for
four }$x${\protect\small \ parameters.}}
\label{fig1}
\end{figure}
The last two terms contain the momentum-dependence of the single-particle
potential. The momentum dependence of the symmetry potential stems from the
different interaction strength parameters $C_{\tau ,\tau ^{\prime }}$ and $%
C_{\tau ,\tau }$ for a nucleon of isospin $\tau $ interacting, respectively,
with unlike and like nucleons in the background fields. More specifically,
we use $C_{unlike}=-103.4$ MeV and $C_{like}=-11.7$ MeV. As an example,
shown in Fig.\ 1 is the density dependence of the symmetry energy 
for $x=-2$, $-1$, $0$ and $1$. 
\begin{figure}[tbh]
\vspace*{-0.8cm} \insertplot{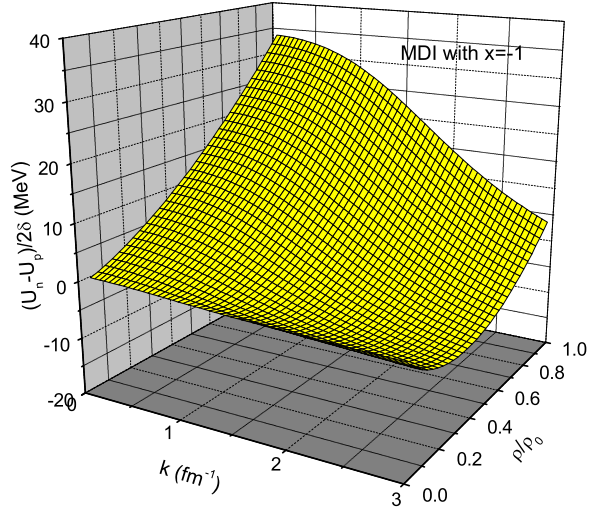} \vspace*{-1cm}
\caption{{\protect\small Symmetry potential as a function of momentum and
density for MDI interaction with }$x=-1${\protect\small .}}
\label{fig2}
\end{figure}

Systematic analyses of a large number of nucleon-nucleus and (p,n) charge
exchange scattering experiments at beam energies below about 100 MeV
indicate undoubtedly that the symmetry potential at $\rho _{0}$, i.e., the
Lane potential, decreases approximately linearly with increasing beam energy 
$E_{kin}$ according to $U_{Lane}=a-bE_{kin}$ where $a\simeq 22-34$ MeV and $%
b\simeq 0.1-0.2$\cite{data1,data2}. This provides a stringent
constraint on the symmetry potential. The potential in eq.\ref{mdi}
meets this requirement very well as seen in Fig.\ 2 where the symmetry
potential $(U_{n}-U_{p})/2\delta $ as a function of momentum and
density for the parameter $x=-1$ is displayed.

One characteristic feature of the momentum dependence of the symmetry 
potential is the different effective masses for neutrons and protons 
in isospin asymmetric nuclear matter, i.e., 
\begin{equation}
\frac{m_{\tau }^{\ast }}{m_{\tau }}=\left\{ 1+\frac{m_{\tau }}{\hbar ^{2}k}%
\frac{dU_{\tau }}{dk}\right\} _{k=k_{\tau }^{F}}^{-1},  \label{mstar}
\end{equation}%
where $k_{\tau }^{F}$ is the nucleon Fermi wave number. With the potential
in eq. \ref{mdi}, we found that the neutron effective mass is higher than
the proton effective mass and the splitting between them increases with both
the density and isospin asymmetry of the medium\cite{lidas03}.

Since both the incoming current in the initial state and the level density
of the final state in nucleon-nucleon (NN) scatterings depend on the
effective masses of colliding nucleons in medium, the in-medium
nucleon-nucleon cross sections are expected to be reduced by a factor $%
\sigma _{NN}^{medium}/\sigma _{NN}=(\mu _{NN}^{\ast }/\mu _{NN})^{2}$, where 
$\mu _{NN}$ and $\mu _{NN}^{\ast }$ are the reduced mass of the colliding
nucleon pairs in free-space and in medium, respectively\cite{sigma}. 
Thus, because of the reduced in-medium nucleon effective masses and their 
dependence on the density and isospin asymmetry of the medium, the 
in-medium NN cross sections are not only reduced compared to their 
free-space values, the nn and pp cross sections are split and the 
difference between them grows in more asymmetric matter. We expect the 
isospin-dependence of the in-medium NN cross sections to play an 
important role in nuclear reactions induced by neutron-rich nuclei\cite{li05}. 

\section{Probing the symmetry energy at subnormal densities with isospin
diffusion}\label{diffusion} 
Tsang et al.\cite{betty04} recently studied the degree of isospin
diffusion in the reaction $^{124}$Sn + $^{112}$Sn by measuring\cite{gsi} 
\begin{equation}
R_{i}=\frac{2X_{^{124}{Sn}+^{112}{Sn}}-X_{^{124}{Sn}+^{124}{Sn}}-X_{^{112}{Sn%
}+^{112}{Sn}}}{X_{^{124}{Sn}+^{124}{Sn}}-X_{^{112}{Sn}+^{112}{Sn}}}
\label{Ri}
\end{equation}%
where $X$ is the average isospin asymmetry $\left\langle \delta
\right\rangle $ of the $^{124}$Sn-like residue. The data is indicated in
Fig. 3 together with the time evolutions of $R_{i}$ and the average 
central densities calculated with $x=-1$ using both the MDI and the 
soft Bertsch-Das Gupta-Kruse (SBKD) interactions are also shown. It 
is seen that the isospin diffusion process occurs mainly from about 
$30$ fm/c to $80$ fm/c corresponding to an average central
density from about $1.2\rho _{0}$ to $0.3\rho _{0}$. The experimental data
from MSU are seen to be reproduced nicely by the MDI interaction with 
$x=-1$, while the SBKD interaction with $x=-1$ leads to a
significantly lower $R_{i}$ value\cite{chen05}. 
\begin{figure}[th]
\insertplot{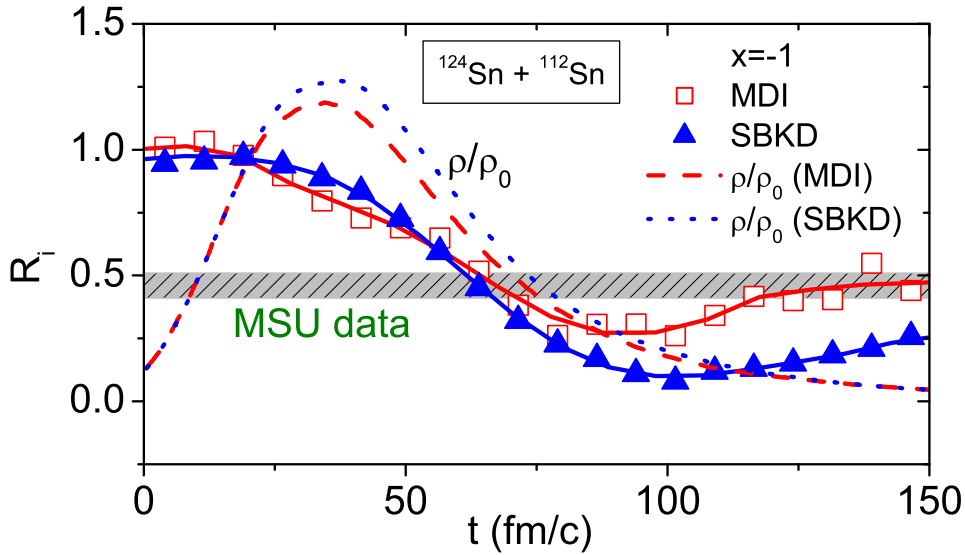}
\caption{{\protect\small The degree of isospin diffusion as a function of
time with the MDI and SBKD interactions. The corresponding evolutions of
central density are also shown.}}
\label{figure3}
\end{figure}

Effects of the symmetry energy on isospin diffusion were also studied by
varying the parameter $x$\cite{chen05}. Only with the parameter $x=-1$ the
data can be well reproduced. The corresponding symmetry energy can be
parameterized as $E_{sym}(\rho )\approx 31.6(\rho /\rho _{0})^{1.05}$. 
In the present study on isospin diffusion, only the free-space NN cross 
sections are used and thus effects completely due to the different 
density dependence of symmetry energy are investigated. As the next step 
we are currently investigating effects of the in-medium NN cross sections 
on the isospin diffusion.

\section{Probing the symmetry energy at supranormal densities}

Several probes that are sensitive to the high density behavior of 
the symmetry energy have been proposed. As an illustration, we 
present here three most sensitive observables that can be measured in 
future experiments at RIA and GSI.

\subsection{Pions yields and $\protect\pi ^{-}/\protect\pi ^{+}$ ratio at
RIA and GSI}\label{RIA1} 

At the highest beam energy at RIA, pion production is
significant. Pions may thus carry interesting information about the EOS of
dense neutron-rich matter\cite{lipi,gaopi}. Shown in Fig.4 are the $\pi ^{-}$
and $\pi ^{+}$ yields as a function of the $x$ parameter. 
\begin{figure}[th]
\insertplot{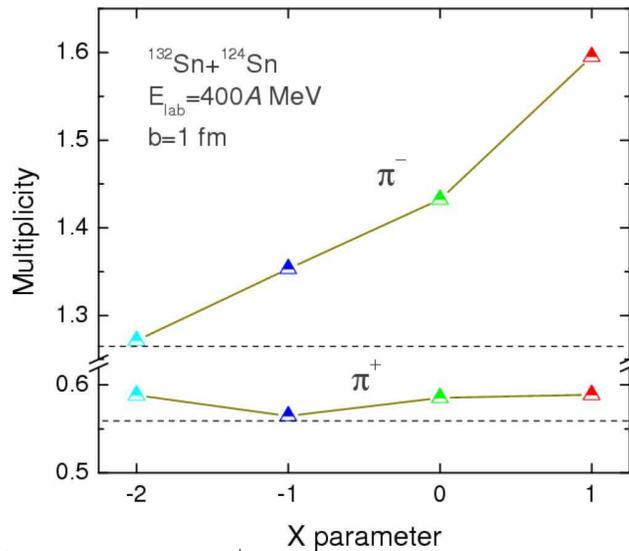}\vspace*{-1cm}
\caption{{\protect\small The $\protect\pi ^{-}$ and $\protect\pi ^{+}$
yields as functions of the $x$ parameter.}}
\label{figure4}
\end{figure}
It is interesting to see that the $\pi ^{-}$ multiplicity depends more
sensitively on the symmetry energy. The $\pi ^{-}$ multiplicity increases 
by about 20\% while the $\pi ^{+}$ multiplicity remains about the same
when the $x$ parameter is changed from -2 to 1. The multiplicity of 
$\pi ^{-}$ is about 2 to 3 times that of $\pi ^{+}$. This is because 
the $\pi ^{-}$ mesons are mostly produced from neutron-neutron 
collisions. Moreover, with the softer symmetry energy the high 
density region is more neutron-rich due to isospin 
fractionation\cite{gaopi}. The $\pi ^{-}$ mesons are thus more
sensitive to the isospin asymmetry of the reaction system and the symmetry
energy. However, one should notice that it is well known that pion yields
are also sensitive to the symmetric part of the nuclear EOS. It is thus hard
to get reliable information about the symmetry energy from $\pi ^{-}$ yields
alone. Fortunately, the $\pi ^{-}/\pi ^{+}$ ratio is a better probe since
statistically this ratio is only sensitive to the difference in the chemical
potentials for neutrons and protons\cite{bertsch}. This expectation is well
demonstrated in Fig.\ 5. It is seen that the pion ratio is quite sensitive
to the symmetry energy, especially at low transverse momenta. Thus, it is
promising that the high density behavior of nuclear symmetry energy $%
E_{sym}(\rho )$ can be probed using the $\pi ^{-}/\pi ^{+}$ ratio. 
\begin{figure}[th]
\insertplot{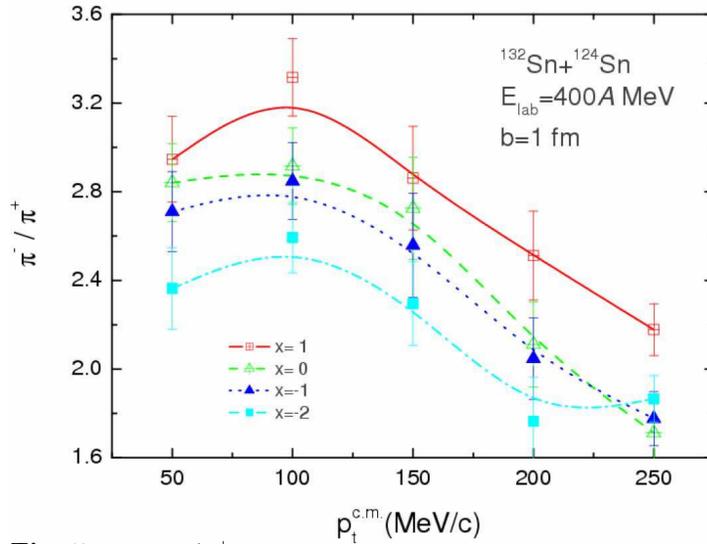}\vspace*{-1cm}
\caption{{\protect\small The $\protect\pi ^{-}/\protect\pi ^{+}$ ratio as a
function of transverse momentum.}}
\label{figure5}
\end{figure}

\subsection{Isospin fractionation and n-p differential flow at RIA and GSI}
\label{RIA2} 

The degree of isospin equilibration or translucency can be
measured by the rapidity distribution of nucleon isospin asymmetry $\delta
_{free}\equiv (N_{n}-N_{p})/(N_{n}+N_{p})$ where $N_{n}$ ($N_{p}$) is the
multiplicity of free neutrons (protons)\cite{li04a}. Although it might be
difficult to measure directly $\delta _{free}$ because it requires the
detection of neutrons, similar information can be extracted from ratios of
light clusters, such as, $^{3}H/^{3}He$, as demonstrated recently within a
coalescence model\cite{chen03b,chen04a}. Shown in Fig.\ 6 are the rapidity
distributions of $\delta _{free}$ with (upper window) and without (lower
window) the Coulomb potential. It is interesting to see that the $\delta
_{free}$ at midrapidity is particularly sensitive to the symmetry energy. As
the parameter $x$ increases from $-2$ to $1$ the $\delta _{free}$ at
midrapidity decreases by about a factor of 2. Moreover, the forward-backward
asymmetric rapidity distributions of $\delta _{free}$ with all four $x$
parameters indicates the apparent nuclear translucency during the reaction%
\cite{ligao}. 
\begin{figure}[th]
\insertplot{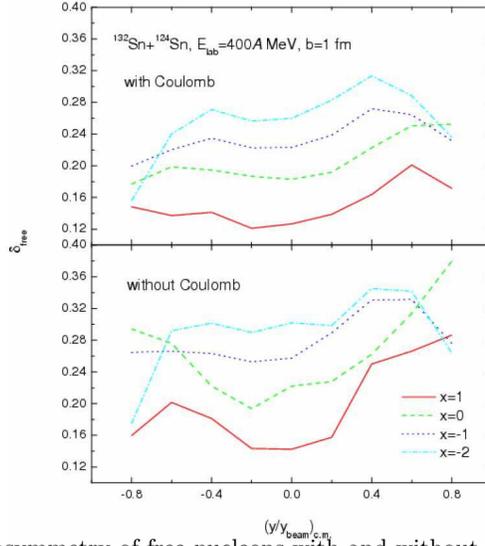}\vspace*{-1cm}
\caption{{\protect\small Isospin asymmetry of free nucleons with and without
the Coulomb force.}}
\label{figure6}
\end{figure}

Another observable that is sensitive to the high density behavior of 
symmetry energy is the neutron-proton differential flow\cite{li00} 
\begin{equation}
F_{n-p}^{x}(y)\equiv \sum_{i=1}^{N(y)}(p_{i}^{x}w_{i})/N(y),
\end{equation}%
where $w_{i}=1(-1)$ for neutrons (protons) and $N(y)$ is the total number of
free nucleons at rapidity $y$. The differential flow combines constructively
effects of the symmetry potential on the isospin fractionation and the
collective flow. It has the advantage of maximizing the effects of the symmetry
potential while minimizing those of the isoscalar potential. Shown in Fig.
7 is the n-p differential flow for the reaction of $^{132}Sn+^{124}Sn$ at a
beam energy of 400 MeV/nucleon and an impact parameter of 5 fm. Effects of
the symmetry energy are clearly revealed by changing the parameter $x$. 
\begin{figure}[th]
\insertplot{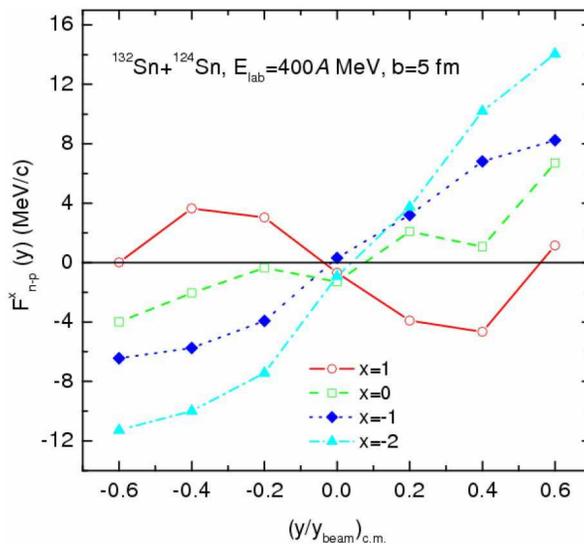}\vspace*{-1cm}
\caption{{\protect\small Neutron-proton differential flow at RIA and GSI
energies}}
\label{figure7}
\end{figure}

\section{Conclusions}
The density dependence of the symmetry energy is very important for 
both nuclear physics and astrophysics. Significant progress has been
made recently by the heavy-ion community in determining the density
dependence of the nuclear symmetry energy. Based on transport model 
calculations, a number of sensitive probes of the symmetry energy have 
been found. The momentum dependence in both the isoscalar and
isovector parts of the nuclear potential was found to play an
important role in extracting accurately the density dependence of the
symmetry energy. Comparing with recent experimental data on 
isospin diffusion from NSCL/MSU, we have extracted a symmetry energy 
of $E_{sym}(\rho )\approx 31.6(\rho /\rho _{0})^{1.05}$ at subnormal 
densities. It would be interesting to compare this conclusion with
those extracted from studying other observables. More experimental
data including neutrons with neutron-rich beams in a broad energy 
range are needed. Looking forward to experiments at RIA and 
GSI with high energy radioactive beams, we hope to pin down 
the symmetry energy at supranormal densities in the near future.

This work was supported in part by the US National Science Foundation of the
under grant No. PHY 0098805, PHYS-0243571 and PHYS0354572, Welch Foundation
grant No. A-1358, and the NASA-Arkansas Space Grants Consortium award
ASU15154. L.W. Chen was supported by the National Natural Science Foundation
of China grant No. 10105008. The work of G.C. Yong and W. Zuo was supported
in part by the Chinese Academy of Science Knowledge Innovation Project
(KECK2-SW-N02), Major State Basic Research Development Program
(G2000077400), the National Natural Science Foundation of China (10235030)
and the Important Pare-Research Project (2002CAB00200) of the Chinese
Ministry of Science and Technology.

\vfill\eject


\begin{thebibliography}{99}
\bibitem{lat01} {\small J.M. Lattimer and M. Prakash, Phys. Rep., \textbf{333%
}, 121 (2000); Astr. Phys. Jour. \textbf{550}, 426 (2001); Science Vol. 
\textbf{304}, 536 (2004); A. W. Steiner, M. Prakash, J.M. Lattimer and P.J.
Ellis, nucl-th/0410066, Phys. Rep. (2005) in press.}

\bibitem{brown} {\small B.A. Brown, Phys. Rev. Let. \textbf{85}, 5296 (2000).%
}

\bibitem{stone} {\small J.R. Stone at al., Phys. Rev. C\textbf{68}, 034324
(2004).}

\bibitem{ireview98} {\small B.A. Li, C.M. Ko and W. Bauer, topical review,
Int. J. Mod. Phys. E\textbf{7}, 147 (1998).}

\bibitem{ibook01} {\small \textit{Isospin Physics in Heavy-Ion Collisions at
Intermediate Energies}, Eds. B. A. Li and W. Uuo Schr\"{o}dear (Nova Science
Publishers, Inc, New York, 2001).}

\bibitem{dan02} {\small P. Danielewicz, R. Lacey and W.G. Lynch, Science
298, 1592 (2002).}

\bibitem{ditoro} {\small V. Baran, M. Colonna, V. Greco and M. Di Toro, 
Phys. Rep. (2005) in press.}

\bibitem{bom1} {\small I. Bombaci, Chapter 2 in ref.\cite{ibook01}.}

\bibitem{betty04} {\small M.B. Tsang et al., Phys. Rev. Lett. \textbf{92},
062701 (2004).}

\bibitem{chen05} {\small L.W. Chen, C.M. Ko and B.A. Li, Phys. Rev. Lett. 
\textbf{94}, 32701 (2005).}

\bibitem{lidas03} {\small B.A. Li, C.B. Das, S. Das Gupta and C. Gale, Phys.
Rev. C\textbf{69}, 011603 (2004); Nucl. Phys. A\textbf{735}, 563 (2004).}

\bibitem{das03} {\small C.B. Das, S. Das Gupta, C. Gale and B.A. Li, Phys.
Rev. C\textbf{67}, 034611 (2003).}

\bibitem{data1} {\small P.E. Hodgson, The Nucleon Optical Model, 1994 (World
Scientific).}

\bibitem{data2} {\small G.W. Hoffmann and W.R. Coker, Phys. Rev. Lett. 
\textbf{29}, 227 (1972).}

\bibitem{sigma} {\small V.R. Pandharipande and S.C. Pieper, Phys. Rev. C%
\textbf{45}, 791 (1991); M. Kohno, M. Higashi, Y. Watanabe, and M. Kawai,
Phys. Rev. C\textbf{57}, 3495 (1998); D. Persram and C. Gale, Phys. Rev. C%
\textbf{65}, 064611 (2002).}

\bibitem{li05} {\small B.A. Li, P. Danielewicz and W.G. Lynch,
nucl-th/0503038.}

\bibitem{gsi} {\small F. Rami et al., Phys. Rev. Lett. \textbf{84}, 1120
(2000).}

\bibitem{lipi} {\small B.A. Li, Phys. Rev. Lett. {88}, 192701 (2002); Nucl.
Phys. A\textbf{708}, 365 (2003); Phys. Rev. C\textbf{67}, 017601 (2003).}

\bibitem{gaopi} {\small B.A. Li, G.C. Yong and W. Zuo, Phys. Rev. C\textbf{71%
}, 014608 (2005).}

\bibitem{bertsch} {\small G.F. Bertsch, Nature \textbf{283}, 280 (1980); A.
Bonasera and G.F. Bertsch, Phys. Let. \textbf{B195}, 521 (1987).}

\bibitem{li04a} {\small B.A. Li, C.M. Ko, and Z.Z. Ren, Phys. Rev. Let. 
\textbf{78}, 1644 (1997); B. A. Li, Phys. Rev. C\textbf{69}, 034614 (2004).}

\bibitem{chen03b} {\small L.W. Chen, C.M. Ko and B.A. Li, Phys. Rev. C%
\textbf{68}, 017601 (2003); Nucl. Phys. \textbf{A729}, 809 (2003).}

\bibitem{chen04a} {\small L.W. Chen, C.M. Ko and B.A. Li, Phys. Rev. C%
\textbf{69}, 054606 (2004).}

\bibitem{ligao} {\small B.A. Li, G.C. Yong and W. Zuo, nucl-th/0412081,
Phys. Rev. C (2005) in press.}

\bibitem{li00} {\small B.A. Li, Phys. Rev. Let. \textbf{85}, 4221 (2000).}
\end{thebibliography}
\end{document}